\documentclass[preprint,superscriptaddress,pra,showpacs,showkeys]{revtex4}
\usepackage{graphicx,amsmath,amsfonts,amssymb}
\usepackage{times}

\begin{document}

\title{Entanglement dynamics and Bell Violations of two atoms in Tavis-Cummings model with phase decoherence
 \footnote{Supported
by the Key Higher Education Programme of Hubei Province under Grant
No Z20052201, the Natural Science Foundation of Hubei Province,
China under Grant No 2006ABA055, and the Postgraduate Programme of
Hubei Normal University under Grant No 2007D20.}}

\author{C. J. Shan\footnote{ E-mail: scj1122@163.com}}
\author{W. W. Cheng}
\author{T. K. Liu\footnote{Corresponding author. E-mail:
tkliuhs@163.com}}
\affiliation{College of Physics and Electronic
Science, Hubei Normal University, Huangshi 435002, China}

\author{D. J. Guo}
\author{Y. J. Xia}
\affiliation{Department of Physics, Qufu Normal University, Qufu
273165, China}
\date{\today}

\begin{abstract}

Considering the dipole-dipole coupling intensity between two atoms
and  the field in the Fock state, the entanglement dynamics between
two atoms that are initially entangled in the system of two
two-level atoms coupled to a single mode cavity in the presence of
phase decoherence has been investigated.  The two-atom entanglement
appears with periodicity without considering phase decoherence,
however, the phase decoherence causes the decay of entanglement
between two atoms, with the increasing of the phase decoherence
coefficient, the entanglement will quickly become a constant value,
which is affected by the two-atom initial state, Meanwhile the
two-atom quantum state will forever stay in the maximal entangled
state when the initial state is proper even in the presence of phase
decoherence. On the other hand, the Bell violation and the
entanglement does not satisfy the monotonous relation, a large Bell
violation implies the presence of a large amount of entanglement
under certain conditions, while a large Bell violation corresponding
to a little amount of entanglement in certain situations. However,
the violation of Bell-CHSH inequality can reach the maximal value if
two atoms are in the maximal entangled state, or vice versa.
\end{abstract}



\maketitle

\section{\textbf{Introduction}}
  Quantum entanglement is one of the most striking features of quantum
mechanics, and plays an important role in  quantum information
processing, such as quantum teleportation[1], quantum dense
coding[2], quantum cryptography[3] and quantum computation[4].
Therefore quantum entanglement has been viewed as an essential
resource for quantum information process, and a great deal of
effort has been devoted to study and characterize the
entanglement. Cavity quantum electrodynamics (QED) techniques has
been recognized as a promising candidate for the physical
realization of quantum information processing. Quantum
entanglement based cavity QED was generated by sending two atoms
being  present simultaneously in the cavity [5] or the two atoms
interacting consecutively with the cavity [6]. However, the above
preparation processes are considered in closed system and the
influences of environment are neglected. Time evolution of
isolated quantum systems is followed by the Schrodinger equation.
But a quantum system  unavoidably interacts with the environment.
The decoherence effect of this interaction will lead to the
degradation of quantum coherence and entanglement. The entangled
state will loss purity and become mixed. Entanglement dynamics
behavior of a quantum system coupled to its environment can
reflect the details of the decoherence effect[7,8]. On the other
hand, entanglement can exhibit the nature of a nonlocal
correlation between quantum systems. Bell's theorem[9] provides a
effective way to test quantum nonlocality[10], quantum nonlocality
will be exhibited if Bell-type inequality is violated for a given
quantum state. Namely, a violation of any Bell-type inequality
gives a quantitative confirmation that a state behaves quantum
nonlocality.
\\
\indent In the original papers, researchers investigated the
entanglement in the JCM[11],  a damped JCM[12] and two-atom
Tavis-Cummings model[13].  Recently Hein etal.[14] investigate
entanglement properties of multipartite states under the influence
of decoherence. Reference [7] shows that quantum mechanical
entanglement can prevail in noisy open quantum systems at high
temperature and far from thermodynamical equilibrium, despite the
deteriorating effect of decoherence. Reference [8]considers the
interaction of a single two-level atom with one of two coupled
microwave cavities and shows analytically that the atom-cavity
entanglement increases with cavity leakage.We investigate the
entanglement time evolution of two entangled two-level atoms that
interact resonantly with a single-mode field in the Fock state[15].
In Ref.[16], the author investigated two two-level atoms coupled to
a single mode optical cavity with the phase decoherence and showed
the rich dynamical features of entanglement arising between atoms
and cavity or between two atoms, however the two-atom dipole-dipole
coupling intensity is neglected, the two atoms are initially in a
separate state and the cavity field is initially prepared in the
vacuum state. In order to study explicitly the entanglement dynamics
of the two-atom system, therefore, in this paper we investigate the
entanglement dynamics between two atoms that are initially in
entangled state in Tavis-Cummings model introducing dipole-dipole
coupling intensity and the field in the Fock state with phase
decoherence, to our knowledge, which has not been reported so far.
In addition quantum nonlocality has been widely studied for the
two-atom entanglement system using Bell-CHSH inequality. Our studies
show that the entanglement between two atoms and Bell-CHSH
inequality decay with phase decoherence and disappear in a constant,
which is affected by two-atom initial state and dipole-dipole
coupling intensity. Meanwhile many new interesting phenomena are
exhibited, e.g., the two-atom quantum state will forever stay in the
maximal entangled state when the initial state is proper even in the
presence of phase decoherence. These interesting phenomena result
from two-atom initial state and dipole-dipole coupling intensity.
The phase decoherence can be used to play a constructive role and
generate the controllable stable  entanglement by adjusting two-atom initial state and dipole-dipole coupling intensity.\\
\indent This paper is organized as follows. We introduce the model
and calculate the reduced density matrices of two two-level atoms in
the next section. In Sec. 3, Entanglement dynamics of two atoms with
phase decoherence have been studied. Sec. 4 gives the relations
between entanglement and Bell violations, and Sec. 5 is the
conclusions.
\section{The model and reduced density matrices of two two-level atoms}
\indent Consider two two-level atoms interacting resonantly with a
single-mode cavity field initially prepared in the Fock state. In
the rotating-wave approximation the Hamiltonian of the atom-field
system reads
\begin{equation}
H=\omega_{0}\sum_{j=1}^{2}S_{j}^{z}+\omega_{a}a^{\dag}a+\sum_{j=1}
^{2}g(a^{\dag}S_{j}^{-}+aS_{j}^{+})+\sum_{i,j=1;{i}\neq{j}}^{2}\Omega
S_{i}^{+} S_{j}^{-}
\end{equation}
where $a$ ($a^{\dagger}$) denotes the annihilation (creation)
operator of the resonant single-mode field, $\omega_{0}$,
$\omega_{a}$  are atomic transition frequency, cavity frequency,
 respectively, $g$ is the coupling
constant between atoms and cavity, $S_{j}^{+}=|e\rangle_{j}\langle
g|$, $S_{j}^{-}=|g\rangle_{j}\langle e|$, $
S_{j}^{z}=\frac{1}{2}(|e\rangle_{j}\langle e|-|g\rangle_{j}\langle
g|)$ are atomic operators, and $\Omega$ is atomic dipole-dipole
coupling constant. In this paper, we investigate the entanglement
between  two atoms by only considering the
 phase decoherence. In this situation, the master
equation governing the time evolution of the system under the
Markovian approximation is given by[17]
\begin{equation}
\frac{d\rho}{dt}=-i[H,\rho]-\frac{\gamma}{2}[H,[H,\rho]]
\end{equation}
where$\gamma$ is the phase decoherence coefficient. The equation
with the similar form has been proposed to describe the intrinsic
decoherence [18]. The formal solution of the master equation (2)
can be expressed as follows [19]:
\begin{equation}
\rho(t)=\sum_{k=0}^{\infty}\frac{(\gamma
t)^{k}}{k!}M^{k}(t)\rho(0)M^{\dagger k}(t)
\end{equation}
where $\rho(0)$ is the density operator of the initial atom-field
system and $M^{k}(t)$ is defined by
\begin{equation}
M^{k}(t)=H^{k}exp(-iHt)exp(-\frac{\gamma t}{2}H^{2})
\end{equation}
We assume $\omega_{0}$ = $\omega_{a}$, the cavity field is prepared
initially in the Fock state $|n\rangle$, atom A and atom B are
prepared in the entangled state
$\cos\theta|eg\rangle+\sin\theta|ge\rangle$, then the initial
density operation for the whole atom-field system is
\begin{equation}
\rho(0)=(\cos\theta|eg\rangle+\sin\theta|ge\rangle)(\cos\theta\langle
eg|+\sin\theta\langle ge|)\otimes|n\rangle\langle n|
\end{equation}
In the subspace of
$K=a^{\dag}a+\frac{1}{2}(S_{1}^{z}+S_{2}^{z})\equiv n$, the
eigenvectors and eigenvalues of Hamiltonian (1) can be written
as[20]
\begin{align}
&|E_{0}\rangle=-\sqrt{\frac{1+n}{1+2n}}|n-1\rangle|ee\rangle+\sqrt{\frac{n}{1+2n}}|n+1\rangle|gg\rangle,
E_{0}=n\omega&\nonumber\\
&|E_{1}\rangle=\frac{1}{\sqrt{2}}(|n\rangle|ge\rangle-|n\rangle|eg\rangle),
E_{1}=n\omega-\Omega&\nonumber\\
&|E_{2}\rangle=\frac{1}{2}\sqrt{\frac{\Delta-\Omega}{\Delta}}(\frac{4\sqrt{n}g}{\Delta-\Omega}|n-1\rangle|ee\rangle-|n\rangle|ge\rangle-|n\rangle|eg\rangle+
\frac{4\sqrt{n+1}g}{\Delta-\Omega}|n+1\rangle|gg\rangle),\nonumber\\
&E_{2}=\frac{1}{2}(2n\omega+\Omega-\Delta)&\\
&|E_{3}\rangle=\frac{1}{2}\sqrt{\frac{\Delta+\Omega}{\Delta}}(\frac{4\sqrt{n}g}{\Delta+\Omega}|n-1\rangle|ee\rangle+|n\rangle|ge\rangle+|n\rangle|eg\rangle+
\frac{4\sqrt{n+1}g}{\Delta+\Omega}|n+1\rangle|gg\rangle),\nonumber\\
& E_{3}=\frac{1}{2}(2n\omega+\Omega+\Delta)&\nonumber
\end{align}
Where $\Delta=\sqrt{8(1+2n)g^{2}+\Omega^{2}}$\\
Substituting $\rho(0)$ into the Eq.(3), the exact time-dependent
density operation can be expressed as
\begin{align}
\rho(t)=&C_{1}|E_{1}\rangle\langle E_{1}|+C_{2}|E_{2}\rangle\langle
E_{2}|+C_{3}|E_{3}\rangle\langle E_{3}|+C_{4}|E_{1}\rangle\langle
E_{2}|+\nonumber\\&C_{5}|E_{2}\rangle\langle
E_{1}|+C_{6}|E_{1}\rangle\langle E_{3}|+C_{7}|E_{3}\rangle\langle
E_{1}|+C_{8}|E_{2}\rangle\langle E_{3}|+C_{9}|E_{3}\rangle\langle
E_{2}|
\end{align}
where
\begin{align}
&C_{1}=\frac{1}{2}(1-\sin2\theta),C_{2}=\frac{1}{4}(1+\sin2\theta)
\frac{\Delta-\Omega}{\Delta},C_{3}=\frac{1}{4}(1+\sin2\theta)
\frac{\Delta+\Omega}{\Delta}&\nonumber\\
&C_{4}=\frac{1}{2\sqrt{2}}\cos2\theta\sqrt{\frac{\Delta-\Omega}{\Delta}}
exp(-\frac{(E_{2}-E_{1})^{2}}{2}\gamma t)exp(i(E_{2}-E_{1})t)&\nonumber\\
&C_{5}=\frac{1}{2\sqrt{2}}\cos2\theta\sqrt{\frac{\Delta-\Omega}{\Delta}}
exp(-\frac{(E_{2}-E_{1})^{2}}{2}\gamma t)exp(-i(E_{2}-E_{1})t)&\nonumber\\
&C_{6}=-\frac{1}{2\sqrt{2}}\cos2\theta\sqrt{\frac{\Delta+\Omega}{\Delta}}
exp(-\frac{(E_{3}-E_{1})^{2}}{2}\gamma
t)exp(i(E_{3}-E_{1})t)&\nonumber\\
&C_{7}=-\frac{1}{2\sqrt{2}}\cos2\theta\sqrt{\frac{\Delta+\Omega}{\Delta}}
exp(-\frac{(E_{3}-E_{1})^{2}}{2}\gamma t)exp(-i(E_{3}-E_{1})t)&\nonumber\\
&C_{8}=\frac{1}{\sqrt{2}}(1+\sin2\theta)\frac{g\sqrt{1+2n}}{\Delta}
exp(-\frac{(E_{3}-E_{2})^{2}}{2}\gamma t)exp(i(E_{3}-E_{2})t)&\nonumber\\
&C_{9}=\frac{1}{\sqrt{2}}(1+\sin2\theta)\frac{g\sqrt{1+2n}}{\Delta}
exp(-\frac{(E_{3}-E_{2})^{2}}{2}\gamma
t)exp(-i(E_{3}-E_{2})t)&\nonumber
\end{align}
The reduced density matrices of the subsystem composed of two
two-level atoms is
\begin{equation}
\rho_{AB}(t)=a_{1}|gg\rangle\langle gg|+a_{2}|ge\rangle\langle ge|+
a_{3}|ge\rangle\langle eg|+a_{4}|eg\rangle\langle
ge|+a_{5}|eg\rangle\langle eg|+a_{6}|ee\rangle\langle ee|
\end{equation}
Where
\begin{align}
a_{1}=(1+\sin2\theta)\frac{2(n+1)g^{2}}{\Delta^{2}}
(1-exp(-\frac{(E_{3}-E_{2})^{2}}{2}\gamma t)\cos(E_{3}-E_{2})t)
\end{align}
\begin{align}
a_{2}=&\frac{1}{2}+(1+\sin2\theta)
\frac{(1+2n)g^{2}}{\Delta^{2}}(-1+
exp(-\frac{(E_{3}-E_{2})^{2}}{2}\gamma
t)\cos(E_{3}-E_{2})t)\nonumber\\& -\frac{1}{4}\cos2\theta
\frac{\Delta-\Omega}{\Delta} exp(-\frac{(E_{2}-E_{1})^{2}}{2}\gamma
t)\cos(E_{2}-E_{1})t\nonumber\\&-\frac{1}{4}\cos2\theta
\frac{\Delta+\Omega}{\Delta} exp(-\frac{(E_{3}-E_{1})^{2}}{2}\gamma
t)\cos(E_{3}-E_{1})t
\end{align}
\begin{align}
a_{3}=a_{4}^{*}=&\frac{\sin2\theta}{2}+(1+\sin2\theta)
\frac{(1+2n)g^{2}}{\Delta^{2}}(-1+exp(-\frac{(E_{3}-E_{2})^{2}}{2}\gamma
t)\cos(E_{3}-E_{2})t )\nonumber\\& -\frac{i}{4}\cos2\theta
\frac{\Delta-\Omega}{\Delta} exp(-\frac{(E_{2}-E_{1})^{2}}{2}\gamma
t)\sin(E_{2}-E_{1})t\nonumber\\&-\frac{i}{4}\cos2\theta
\frac{\Delta+\Omega}{\Delta} exp(-\frac{(E_{3}-E_{1})^{2}}{2}\gamma
t)\sin(E_{3}-E_{1})t
\end{align}
\begin{align}
a_{5}=&\frac{1}{2}+(1+\sin2\theta)
\frac{(1+2n)g^{2}}{\Delta^{2}}(-1+
exp(-\frac{(E_{3}-E_{2})^{2}}{2}\gamma
t)\cos(E_{3}-E_{2})t)\nonumber\\& +\frac{1}{4}\cos2\theta
\frac{\Delta-\Omega}{\Delta} exp(-\frac{(E_{2}-E_{1})^{2}}{2}\gamma
t)\cos(E_{2}-E_{1})t\nonumber\\&+\frac{1}{4}\cos2\theta
\frac{\Delta+\Omega}{\Delta} exp(-\frac{(E_{3}-E_{1})^{2}}{2}\gamma
t)\cos(E_{3}-E_{1})t
\end{align}
\begin{align}
&a_{6}=(1+\sin2\theta)\frac{2n g^{2}}{\Delta^{2}}
(1-exp(-\frac{(E_{3}-E_{2})^{2}}{2}\gamma t)\cos(E_{3}-E_{2})t)&
\end{align}
\section{Entanglement dynamics of two atoms with phase decoherence}
\indent In order to discuss the entanglement dynamics in the above
system, we adopt the negative eigenvalues of the partial
transposition to quantify the degree of entanglement. The idea of
this measure of the entanglement is the Peres-Horodecki criterion
for the separability of bipartite systems [21]. The state is
separable if the partial transposition is a positive operator,
however, if one of the eigenvalues of the partial transposition is
negative then the state is entangled. For a two-qubit system
described by the density operator, the negativity can be defined
by:[22]
\begin{equation}
E_{AB}=-2\sum_{i}\mu_{i}
\end{equation}
where $\mu_{i}$ are the negative eigenvalues of the partial
transposition of $\rho_{AB}^{\Gamma}$. When $E_{AB}=0$, the two
qubits are separable and $E_{AB}=1$  indicates maximal entanglement
between them. \\
\indent We can make a partial transposition for atom B and work out
the eigenvalues of the partial transposition $\rho_{AB}^{\Gamma}$.
The four eigenvalues are $a_{2}$, $a_{5}$,
$\frac{1}{2}(a_{1}+a_{6}+\sqrt{a_{1}^{2}-2a_{1}a_{6}+a_{6}^{2}+4a_{3}a_{4}})$,
$\frac{1}{2}(a_{1}+a_{6}-\sqrt{a_{1}^{2}-2a_{1}a_{6}+a_{6}^{2}+4a_{3}a_{4}})$.
Substitute them into Eq.(14), the explicit expression of $E_{AB}$
characterizing the entanglement of two atoms can be found to be
\begin{align}
&E_{AB}=|a_{2}|+|a_{5}|+|\frac{1}{2}(a_{1}+a_{6}+\sqrt{a_{1}^{2}-2a_{1}a_{6}+a_{6}^{2}+4a_{3}a_{4}})|+
\nonumber\\&|\frac{1}{2}(a_{1}+a_{6}-\sqrt{a_{1}^{2}-2a_{1}a_{6}+a_{6}^{2}+4a_{3}a_{4}})|
-(a_{2}+a_{5}+\frac{1}{2}(a_{1}+a_{6}+\sqrt{a_{1}^{2}-2a_{1}a_{6}+a_{6}^{2}+4a_{3}a_{4}})\nonumber\\&
+\frac{1}{2}(a_{1}+a_{6}-\sqrt{a_{1}^{2}-2a_{1}a_{6}+a_{6}^{2}+4a_{3}a_{4}}))
\end{align}
\indent In the following, we analyze the numerical results for the
time
evolution of the two-atomic entanglement. \\
\indent We firstly consider the case of $\gamma=0$, i.e., the
absence of phase decoherence. The time evolution behaviors of the
entanglement are showed in Fig.1-Fig.3(assuming g=1 in all the
figures in this paper) with different initial state and
dipole-dipole
coupling intensity for $g=1,n=0$.  \\
\indent Figure 1 depicts the time evolution of the entanglement when
the pair of atoms are initially prepared in the different states. It
is observed that the entanglement evolves periodically in the
absence of phase decoherence. We consider three cases of the initial
state, i.e., the disentangled of the two atoms ((a)and(c), solid
line ), not maximal entangled state ((a), (b), (c) and(d), dashed
line) and maximal entangled  state ((b)and(d), solid line). In the
first case, we can observe that the two atoms that are initially
separate can generate entanglement by the atom-field interaction and
atom-atom interaction. At certain time the entanglement evolves to
its zero and the two two-level atoms are disentangled, while at the
large time scale the two atoms are entangled. In a period, the
degree of the entanglement increases gradually to a larger
value(about 0.5), then decreases to a smaller value(about 0.2), then
again increases and finally decreases to zero.  In the second case,
the two atoms own the same  entanglement at $t=0$, but have
different phase angles. It is the phase angle that leads to
considerable different time evolution of the entanglement. One case
is that the degree of the two-atom entanglement is no more than that
of the initial entanglement, as is shown in Fig.1((a)and (b), dashed
line), the other case is that the degree of the two-atom
entanglement is more than that of the initial entanglement all the
time during  the interaction, the peak of the entanglement
increases, as is shown in Fig.1((c)and (d), dashed line), which
means the larger entangled state can be prepared by choosing the
initial phase angle. The third case is that the two atoms are
initially in the maximal entangled state. In Fig.1((c), solid line),
the time evolution is similar to the above case, however, from
Fig.1((d), solid line), we can find the two-atom quantum state will
forever stay in the maximum entangled state when the initial state
is proper, this corresponding to the fact that the two atoms do not
show any dynamic evolution
and remain the initial state.\\
\indent Figure 2 displays the time evolution of the entanglement for
two values of no and weak dipole-dipole interaction. Fig.2((a),
solid line) corresponding to the case of being no  dipole-dipole
interaction, the peak of the maximum entanglement becomes small
comparing with the case of that in  Fig.1((a),solid line,
$\Omega=1$), Since there is no dipole-dipole interaction between the
two atoms, it is very clear that this entanglement is induced purely
by atom-field interaction. This is consistent with Ref. [11]. The
dipole-dipole interaction plays a constructive role in the
entanglement formation between two atoms. From these figures, we can
see that the degree of the entanglement is not necessarily increases
with the increase of dipole-dipole interaction. In Fig.2(c), the
degree of the entanglement can reach the maximum value 1 and the
range of the oscillation becomes larger comparing with the situation
in Fig.1(a), while the value of the dipole-dipole interaction in
Fig.2(c) is less than that in Fig.1(a). It is interesting to find
that the two atoms can generate maximal entangled state even they
are separate initially by
adjusting the dipole-dipole interaction.\\
\indent In Fig.3, we consider the situation of strong dipole-dipole
interaction. With the increase of dipole-dipole interaction, the
period of the oscillation becomes short. The time evolution
character is  similar to the case of the weak dipole-dipole
interaction for the separate initial state. However, for the
entangled initial state, that is not the case. An interesting result
is the entanglement between the two atoms increases to a larger
value than the initial entanglement in Fig.3((a)and (c), dashed
line), while the entanglement decreases in in Fig.1((a), dashed
line). In the strong coupling case, i.e., $\Omega \gg g $, from
Eq.(9-13), we can see that dipole-dipole interaction $\Omega$ plays
a key role in the quantum entanglement between the atom. Atom-atom
interaction reduces the atom-field interaction. That is to say,
strong dipole-dipole interaction is helpful for the entanglement
production.
\\
\indent Let us now turn to discuss the condition of existing phase
decoherence($\gamma\neq0$). The phase decoherence causes the decay
of the entanglement between two atoms, which is shown in
Figs.4(a)and 4(b). With the increase of phase decoherence
coefficient, the initial entanglement oscillates with time and will
gradually become a constant value, which depends on the initial
state of the two atoms. That is to say, the phase decoherence in the
atom-field interaction suppresses the entanglement, but the phase
decoherence can not fully destroy the entanglement between two
atoms. From Figs.4(a)and 4(b), we can also see that the pairwise
entanglement between two atoms can achieve a very large value even
in the presence of phase decoherence, which is similar to the case
without phase decoherence. For the proper initial state, their
entanglement can be preserved during the time evolution as its
initial value with phase decoherence. The above time evolution
character arises due to in the time evolution the additional term in
Eq.(2) leads to the appearance of the decay factor, which are
responsible for the destruction of the entanglement. In order to
discuss how the entanglement changes with the dipole-dipole
interaction, in Figs.4(c)and 4(d) we give the plot of the
entanglement for $\Omega=0.5$ and $\Omega=5$ in the present of
$\gamma=0.1$. The result is that  more stronger the dipole-dipole
interaction is, more faster the entanglement does oscillate. As for
the situation of strong dipole-dipole interaction, the entanglement
decreases rapidly, then approaches to a stable value, which is
different from the case in the absent of phase decoherence. what
affects the stable value? From Eq.(15), it is easy to verify that
$E_{AB}$ in the case of $\gamma\neq 0$ for given long time,
\begin{equation}
E_{AB}=\frac{-2(1+2n)g^{2}(\sin\theta+\cos\theta)^{2}+\sqrt{4g^{4}(\sin\theta+\cos\theta)^{4}+
(1+2n)^{2}(-2g^{2}+6(g^{2}+\Omega^{2})\sin2\theta)^{2}}}
{8(1+2n)g^{2}+\Omega^{2}}
\end{equation}
which means that the entanglement of stationary state depends on the
initial state, the dipole-dipole interaction and the field in the
Fock state. One may question whether there exists a situation in
which two atoms can forever achieve maximal entanglement in the
present of phase decoherence. Fig.4((a)and(b), dash dot line) give
the answer. What is the reason why the two atoms can stay the
maximal entanglement in the present of phase decoherence ? From
Eq.(9-13), we can see $a_{1}=a_{6}=0$, $a_{2}=a_{5}=\frac{1}{2}$,
$a_{3}=a_{4}=-\frac{1}{2}$ if the angles satisfy the following
relation $\theta=\frac{(4k-1)\pi}{4}, k=1, 2, \cdots$. The two atoms
are in the maximally entangled  state
$\frac{1}{\sqrt{2}}(eg\rangle-|ge\rangle)$, so the entanglement has
nothing with the phase decoherence coefficient, the two-atom initial
state, the dipole-dipole coupling intensity between two atoms and
the field in the Fock state.\\
\indent At the end of this section, we discuss to achieve
entanglement between the two atoms if the initial atoms are prepared
in different states and the cavity field is in the Fock state. In
Fig.5, we plot the entanglement  as the function of time t for
different values of phase decoherence rate $\gamma$ and
dipole-dipole coupling intensity $\Omega$ if the field in the
$|1\rangle$. Two cases are shown in Fig.5(a) for different
dipole-dipole coupling intensity if $\gamma=0$, i.e the entanglement
between two atoms of being no  dipole-dipole interaction falls off
 while $E_{AB}$ increases having dipole-dipole interaction as n increases for the  initial separate two-atom
state. The influence is completely different compared to that for
the $n=0$ case.  For the  initial entangled two-atom state, the
notable difference here is that the peak of the entanglement becomes
larger than that in Fig.1, while at some time, the two atoms stay in
the separate state. The photon number n helps to increase the peak
value of entanglement. Figs.5(c) and 5(d) corresponding to the case
of phase decoherence $\gamma=0.1$. An interesting comparison can be
made with the case of the field in the vacuum state. The
entanglement decays sharply as n increases and the stationary state
entanglement is affected by the Fock state, so we can get two-atom
entanglement mediated by the Fock state cavity field.
\\
\indent From the above analysis, it is clear to note that the phase
decoherence coefficient, the two-atom initial state, the
dipole-dipole coupling intensity between two atoms and the field in
the Fock state have notable influence on the entanglement of two
atoms.
\section{Bell violations and the relations between entanglement and Bell violations}
 The quantum nonlocal property can be characterized
by the maximal violation of Bell's inequality. Jeong etal.[23]
have defined the maximal violation of Bell's inequality as
mearurement of the degree of quantum nonlocality. Here we discuss
the CHSH inequality. The CHSH operator is defined by[24]
\begin{equation}
\vec{B}=({\vec{a}}\cdot{\vec{\sigma}})\otimes(\vec{b}\cdot{\vec{\sigma}})
+({\vec{a}}\cdot{\vec{\sigma}})\otimes(\vec{b'}\cdot{\vec{\sigma}})
+ ({\vec{a'}}\cdot{\vec{\sigma}})\otimes(\vec{b}\cdot{\vec{\sigma}})
+({\vec{a'}}\cdot{\vec{\sigma}})\otimes(\vec{b'}\cdot{\vec{\sigma}})
\end{equation}
where $\vec{a}$, $\vec{a'}$, $\vec{b}$, $\vec{b'}$ are unit
vectors. The hidden variable theories impose the Bell-CHSH
inequality $|<\vec{B}>|\leq2$ where $<\vec{B}>$ is the mean value
of the bell operation for a given quantum state. However, in the
quantum theory it is found that $|<\vec{B}>|\leq2\sqrt{2}$, which
implies the Bell-CHSH inequality is violated. The maximal amount
of Bell's violation of a state $\rho$ is given by [25]
\begin{equation}
<B>=2\sqrt{\lambda+\lambda'}
\end{equation}
Where $\lambda$, $\lambda'$ are the two largest eigenvalues of
$T_{\rho}^{\dag}T_{\rho}$, the elements of matrix $T_{\rho}$ are
$(T_{\rho})_{nm}=Tr(\rho\sigma_{n}\otimes\sigma_{m})$, here
$\sigma_{1}=\sigma_{x}$, $\sigma_{2}=\sigma_{y}$, and
$\sigma_{3}=\sigma_{z}$ denote the usual Pauli matrices. For the
density operator in Eq. characterizing the time evolution of two
atoms, $\lambda+\lambda'$ can be written as follows:
\begin{equation}
\lambda+\lambda'=4a_{3}a_{4}+\max[4a_{3}a_{4},(a_{1}+a_{6}-a_{2}-a_{5})^{2}]
\end{equation}
it is easy to draw the violation of Bell¡¯s inequality for two
atoms.
\begin{equation}
<B>=2\sqrt{4a_{3}a_{4}+\max[4a_{3}a_{4},(a_{1}+a_{6}-a_{2}-a_{5})^{2}]}
\end{equation}
\indent Similarly, Figs.6-8 display the numerical results of the
analytical expression of maximal violation of Bell's inequality for
the field in the vacuum state. In Fig.6, we plot the time evolution
of the maximal violation of Bell's inequality for $\Omega=1$ and
$\Omega=0.5$ when the two atoms are prepared in different states.
For the sepatate initial state, our calculations show that two atoms
cannot violate the CHSH inequality in this case, which is seen in
Fig.6((a)dashed line). If we appropriately choose the value of the
dipole-dipole interaction $\Omega$, From Fig.6(c)(dashed line), an
interesting result is that two atoms can violate the CHSH inequality
in certain time. Even the two atoms have the same entanglement and
the phase angle, it is the dipole-dipole interaction that makes the
CHSH inequality of the two atoms evolve in different ways. The
violation of the CHSH inequality increases firstly in Fig.6(b)(solid
line), while the violation the CHSH inequality decreases firstly in
Fig.6(d)(dashed line). In addition the violation of Bell-CHSH
inequality can stay in the maximal value when the entanglement angle
satisfies $\theta=3\pi/4$. Fig.7 corresponding to the time evolution
of Bell-CHSH inequality in the present of phase decoherence. Fig.8
depicts the time evolution of Bell-CHSH inequality against the
strong dipole-dipole interaction with the phase decoherence and
without the phase decoherence. The result is expected as it is shown
in Figs.8(a) and 8(b) that the strong dipole-dipole interaction
maximize the violation of the CHSH inequality, in this case the
larger violation of Bell-CHSH inequality can be achieved. Similar to
the influence of phase decoherence on the entanglement, the
violation of Bell-CHSH inequality is very fragile against the phase
decoherence and finally disappears in the different stationary state
with different initial
state and dipole-dipole coupling intensity. \\
\indent In the following, we are devoted to settling the
relationship between entanglement, measured in terms of the
negativity, and the Bell violations in the system [1]. And although
the quantitative relations have never been investigated in detail,
it is quite often suggested that a large Bell violation implies the
presence of a large amount of entanglement and vice versa. Recently,
Verstraete et al. investigated the relations between the violation
of the CHSH inequality and the concurrence for systems of two
qubits[26]. For the pure states and some Belldiagonal states, the
maximal value of B for given concurrence C is $2\sqrt{1+C^{2}}$. If
the concurrence $C\geq\sqrt{2}/2$, the minimal value of B is
$2\sqrt{2}C$, furthermore, the entangled two-qubits state  may not
violate any CHSH inequality with the concurrence C$\leq\sqrt{2}/2$,
except their Belldiagonal normal form does violate the CHSH
inequalities. Comparing Fig.1((a) solid line) with Fig.5((a) solid
line), we can find that though two atoms get entangled in the time
evolution, two atoms cannot violate the CHSH inequality in this
case. Fig.5 shows two atoms can violate the CHSH inequality in the
case that the entanglement is larger than a certain value. Under
certain condition, the more Bell violation, the larger amount of
entanglement. However, the violation of Bell¡¯s inequality is not a
sufficient condition for the entanglement, that is to say, a large
Bell violation is not necessarily with a large amount of
entanglement, which can be seen in Fig.1((c) dashed line), Fig.2((d)
solid line), Fig.6((b)solid line, (d)dashed line). In Fig.3((a)
solid line) and Fig.8((a) solid line). The dipole-dipole interaction
decreases the degree of violation while increases the amount of
entanglement. One interesting point is that the entanglement degree
is initially very little, while the violation of Bell¡¯s inequality
can be generated, according to Re.[23], we can know the Bell
diagonal normal form in system (1) does violate the CHSH
inequalities. Our calculations also show that the condition of the
maximal violation is that the entanglement degree is  maximal. In a
word, the Bell violation and entanglement does not satisfy the
monotonous relation. This is consist with Re.[13]. So this
phenomenon is still valid for the form of Bell's inequality and the
entanglement measurement in
 this paper.\\
\section{Conclusion}
In summary, we have studied quantum entanglement and quantum
nonlocality of two atoms in Tavis-Cummings model with phase
decoherence. It is shown that  the phase decoherence causes the
decay of entanglement between two atoms. With the increasing of the
phase decoherence coefficient, the entanglement will quickly become
a constant value, which is affected by the two-atom initial state,
the dipole-dipole coupling intensity and the field in the Fock
state. Therefore, the amount of the entanglement can be increased by
adjusting the two-atom initial state, the dipole-dipole coupling
intensity and the field in the Fock state. The violation of
Bell-CHSH inequality is very fragile against the phase decoherence
and finally disappears in the different stationary state in the
absence of phase decoherence. In addition, the relationship between
the entanglement and the nonlocality of two atoms is investigated,
under certain conditions either a larger violation or a less
violation can be generated with the increasing of entanglement. We
hope that the results obtained in this paper would find their
applications in quantum
information  processing and the test of quantum nonlocality.\\

\newpage
\begin{figure}
\begin{center}
\includegraphics[width=1.0\textwidth]{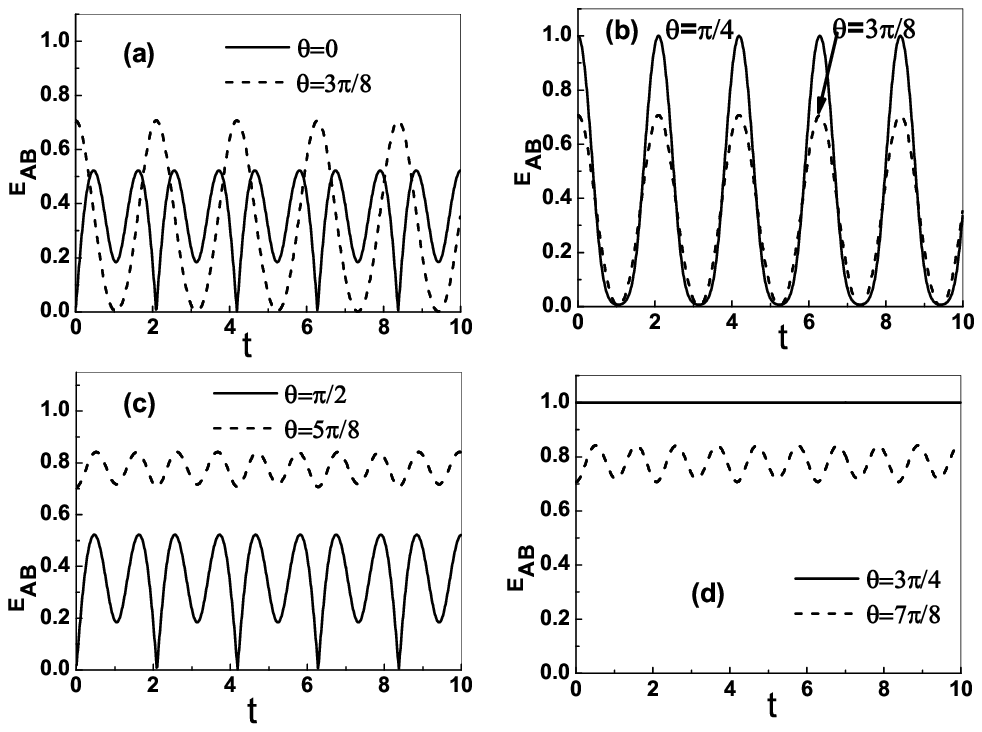}\\
\caption{The entanglement between the two atoms ($E_{AB}$) is
plotted as a function of time t with $g=1,\Omega=1,\gamma=0,n=0$
when the two-atomic state is initially prepared in the different
state}\label{Fig.1.EPS}
\end{center}
\end{figure}

\begin{figure}
\begin{center}
\includegraphics[width=1.0\textwidth]{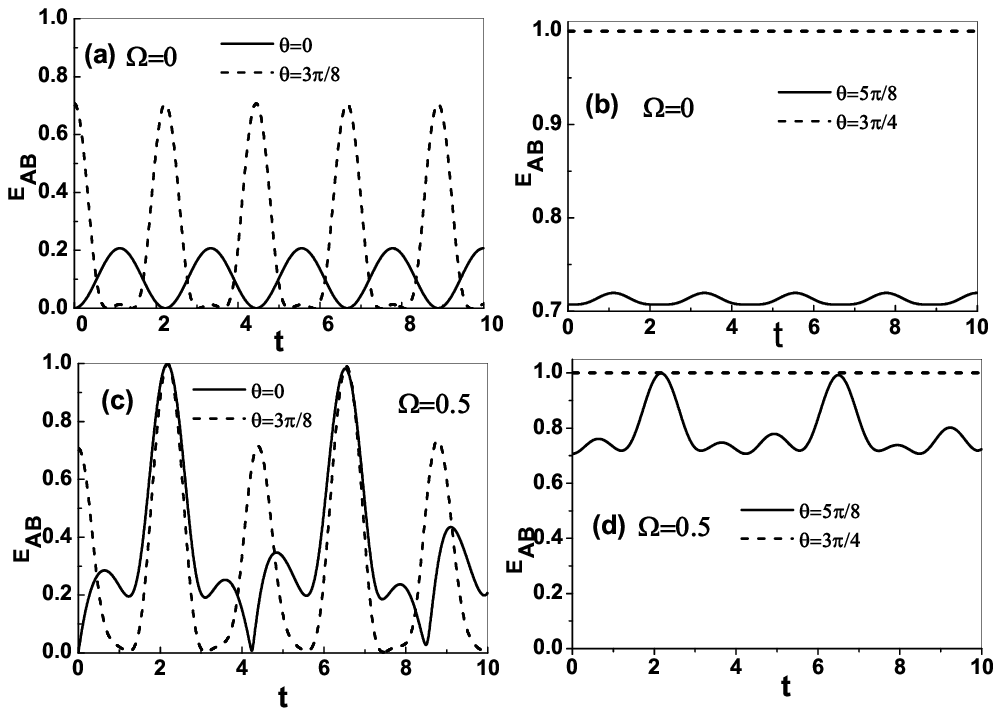}\\
\caption{The entanglement between the two atoms ($E_{AB}$) is
plotted as a function of time t with
$g=1,\gamma=0,n=0$.}\label{Fig.2.EPS}
\end{center}
\end{figure}

\begin{figure}
\begin{center}
\includegraphics[width=1.0\textwidth]{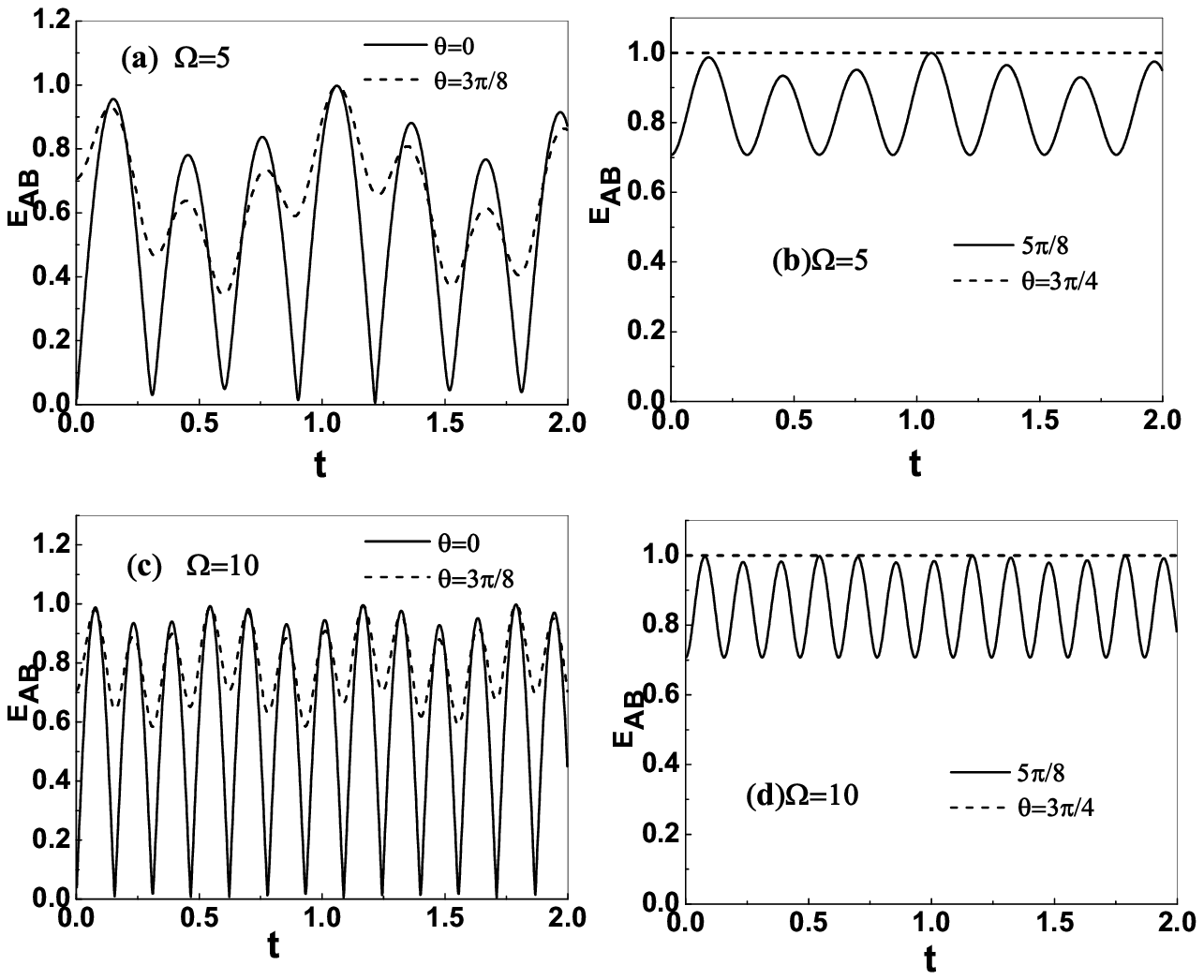}\\
\caption{The entanglement between the two atoms ($E_{AB}$) is
plotted as a function of time t with
$g=1,\gamma=0,n=0$.}\label{Fig.3.EPS}
\end{center}
\end{figure}

\begin{figure}
\begin{center}
\includegraphics[width=1.0\textwidth]{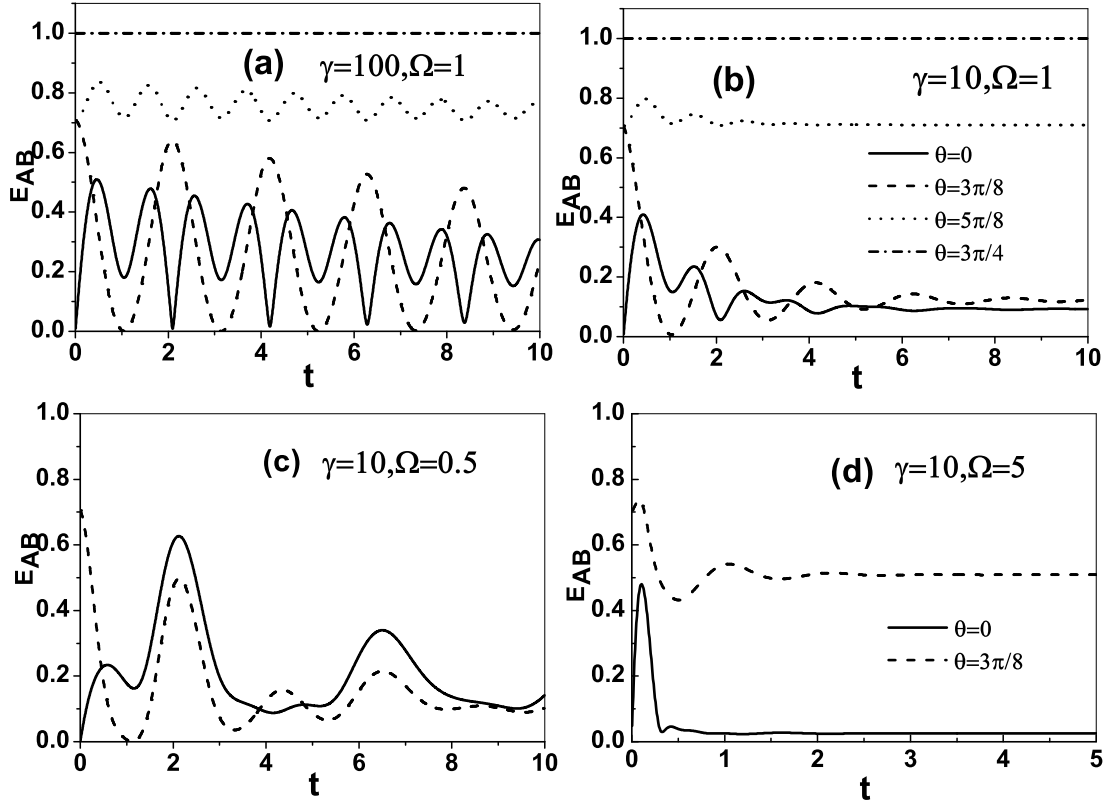}\\
\caption{The entanglement between the two atoms ($E_{AB}$) is
plotted as a function of time t with $g=1,n=0$ in the present of
phase decoherence.}\label{Fig.4.EPS}
\end{center}
\end{figure}

\begin{figure}
\begin{center}
\includegraphics[width=1.0\textwidth]{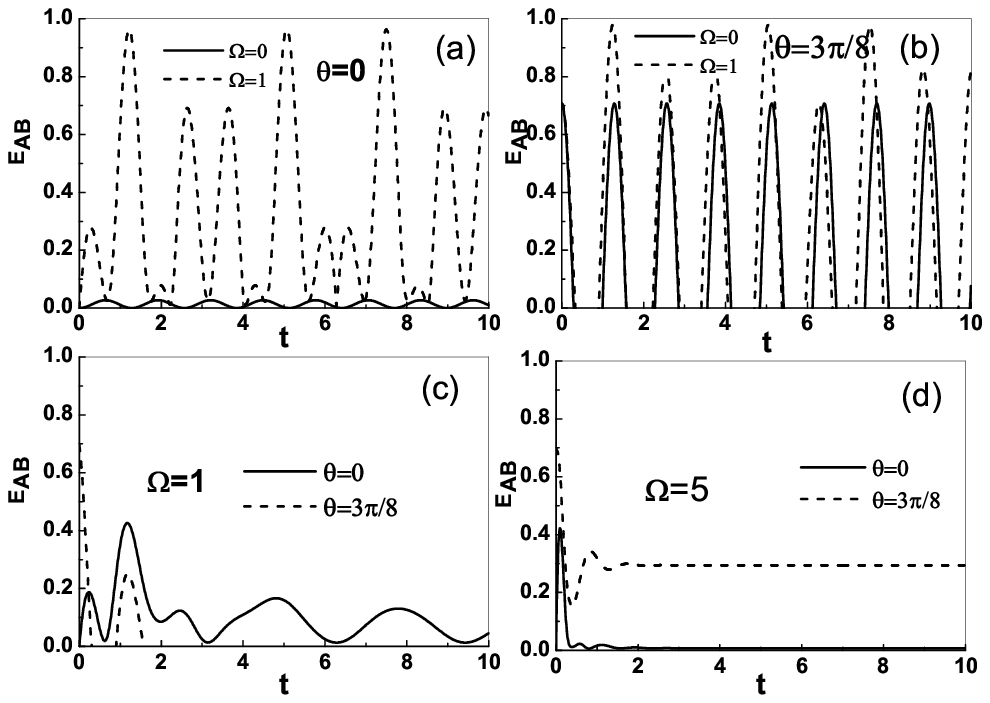}\\
\caption{TThe entanglement between the two atoms ($E_{AB}$) is
plotted as a function of time t with $g=1,n=1$, (a) and (b)$\gamma=0$, while (c) and (d)$\gamma=0.1$ \\
\indent Fig.6 The time evolution of maximal violation of Bell-CHSH
inequality for $g=1,\gamma=0,n=0$.}\label{Fig.5.EPS}
\end{center}
\end{figure}

\begin{figure}
\begin{center}
\includegraphics[width=1.0\textwidth]{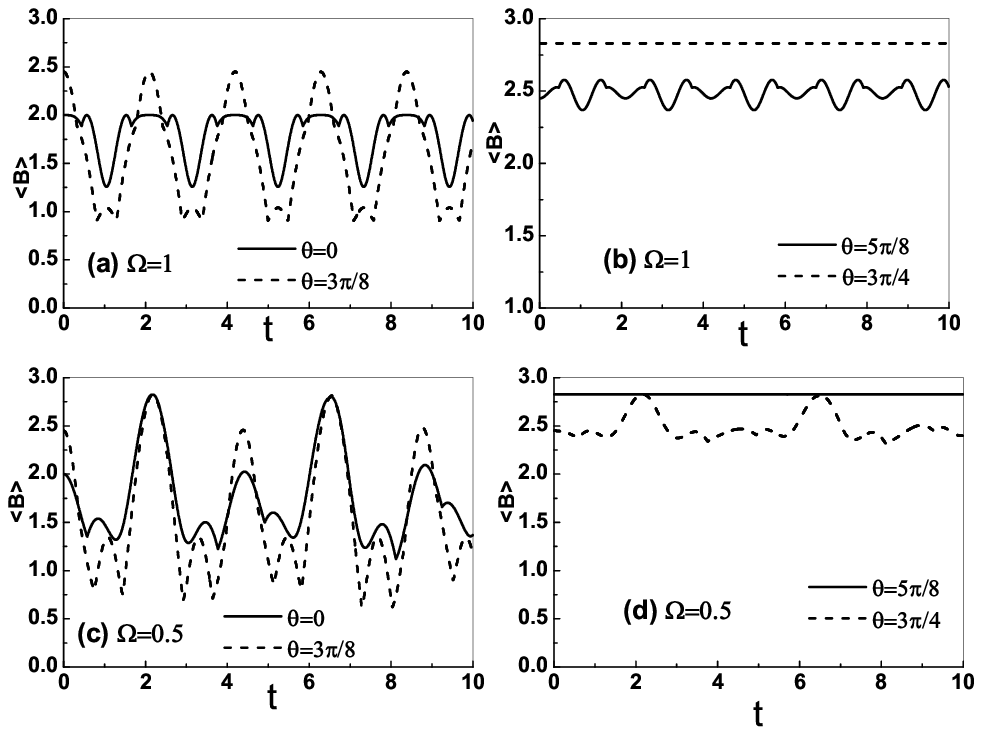}\\
\caption{The time evolution of maximal violation of Bell-CHSH
inequality for $g=1,\gamma=0,n=0$.}\label{Fig.6.EPS}
\end{center}
\end{figure}

\begin{figure}
\begin{center}
\includegraphics[width=1.0\textwidth]{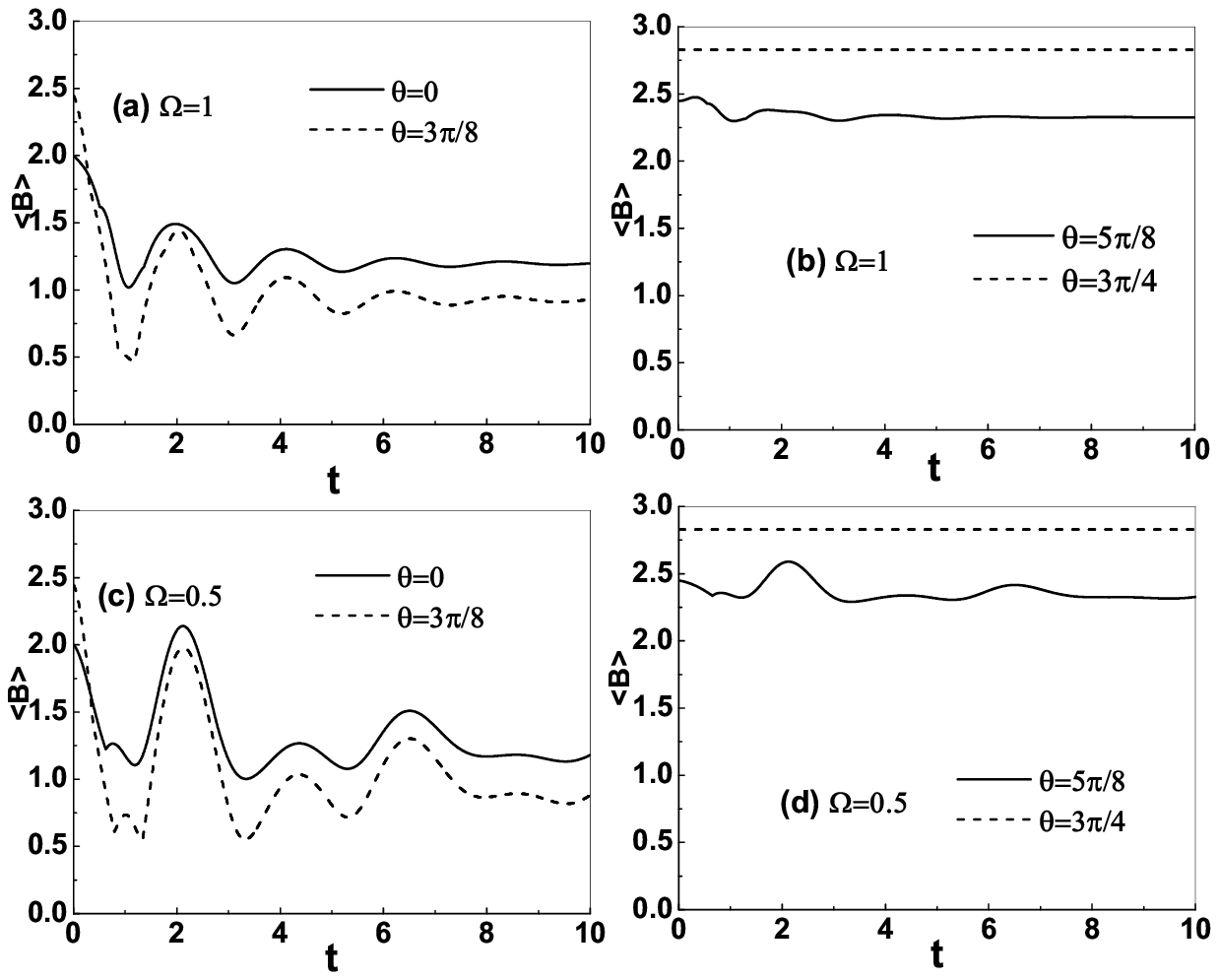}\\
\caption{The time evolution of maximal violation of Bell-CHSH
inequality for $g=1,\gamma=0.1,n=0$.}\label{Fig.7.EPS}
\end{center}
\end{figure}

\begin{figure}
\begin{center}
\includegraphics[width=1.0\textwidth]{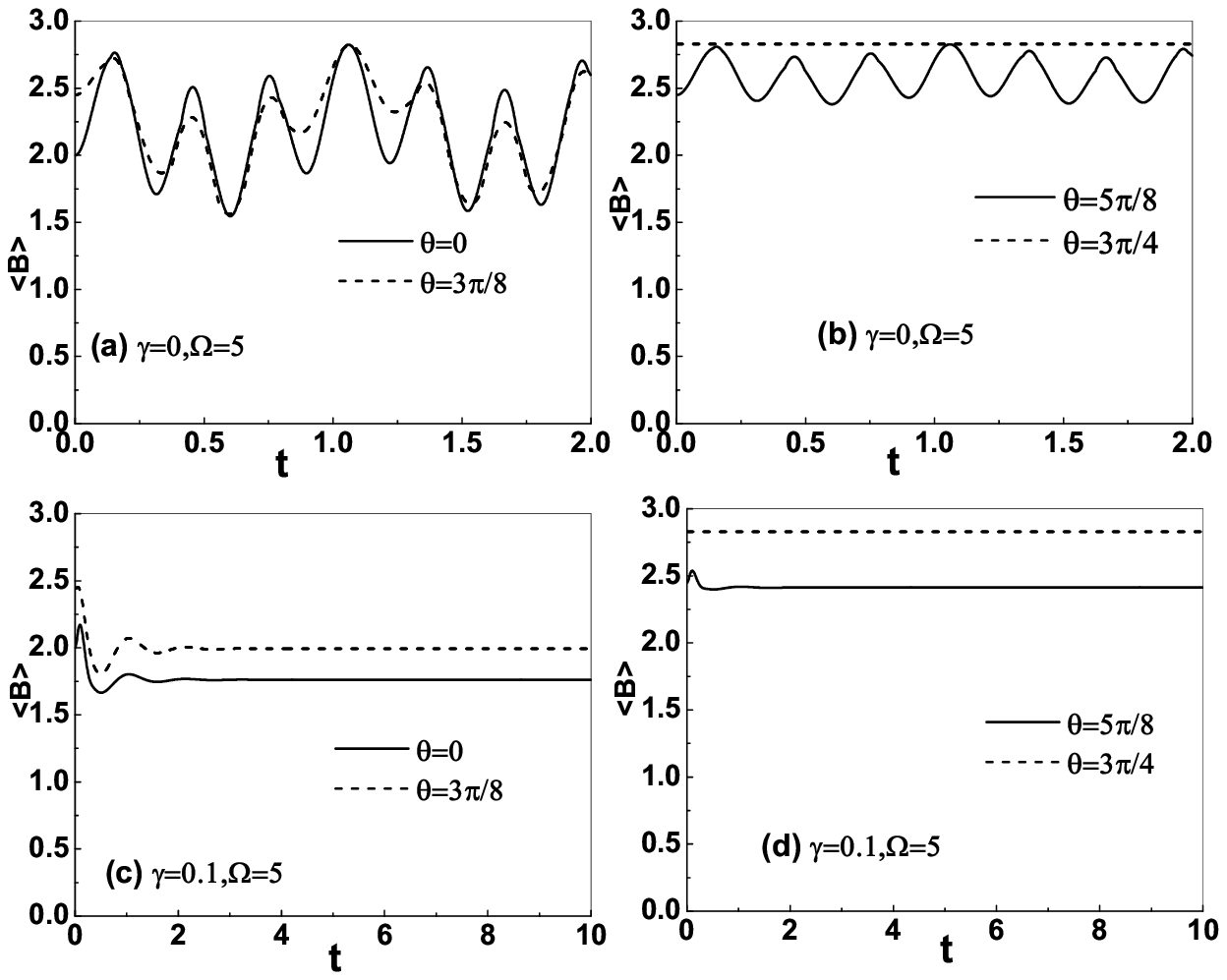}\\
\caption{The time evolution of maximal violation of Bell-CHSH
inequality for $g=1,\Omega=5,n=0$.}\label{Fig.8.EPS}
\end{center}
\end{figure}

\end{document}